\documentclass[authoryear,12pt,letter,3p]{jowarticle}
\usepackage{graphicx}
\usepackage{amsmath}
\usepackage{amssymb}
\usepackage{setspace}
\usepackage[all]{xy}
\makeatletter
\renewcommand\section{\@startsection{section}{1}{\z@}{-3.25ex plus -1ex minus -.2ex}{1.5ex plus .2ex}{\normalsize\bf}}
\renewcommand\subsection{\@startsection{subsection}{2}{\z@}{-3.25ex plus -1ex minus -.2ex}{1.5ex plus .2ex}{\normalsize\bf}}
\renewcommand\subsubsection{\@startsection{subsubsection}{3}{\z@}{-3.25ex plus -1ex minus -.2ex}{1.5ex plus .2ex}{\normalsize\bf}}
\makeatother

\newtheorem{thm}{Theorem}[section]

\numberwithin{equation}{section}

\newcommand{\supp}[1]{\text{supp}(#1)}

\begin{document}
\begin{frontmatter}
\title{Inertial motion, explanation, and the foundations of classical spacetime theories}
\author{James Owen Weatherall}\ead{weatherj@uci.edu}
\address{Department of Logic and Philosophy of Science\\ 3151 Social Science Plaza A, University of California, Irvine, CA 92697}
\begin{abstract}I begin by reviewing some recent work on the status of the geodesic principle in general relativity and the geometrized formulation of Newtonian gravitation.  I then turn to the question of whether either of these theories might be said to ``explain'' inertial motion.  I argue that there is a sense in which \emph{both} theories may be understood to explain inertial motion, but that the sense of ``explain'' is rather different from what one might have expected.  This sense of explanation is connected with a view of theories---I call it the ``puzzleball view''---on which the foundations of a physical theory are best understood as a network of mutually interdependent principles and assumptions.\end{abstract}
\end{frontmatter}

\doublespacing

\section{Introduction}

There is a very old question in the philosophy of space and time, concerning how and why bodies move in the particular way that they do in the absence of any external forces.  The question originates with Aristotle, and indeed, the puzzle is particularly acute when one thinks of it as the ancients might have.  Given some external influence on a body, it might seem clear why that body moves in one fashion rather than another: the external influence forces it to do so.  But when there are no forces present, what does the work of picking one possible state of motion over any other?  Consider planetary motion: there are no apparent forces acting on planets, and yet they proceed along fixed trajectories.  Why these orbits rather than others?  In Aristotelian terms, what determines the ``natural motions'' of a body?

The modern answer to the question originates with Galileo and Descartes, but finds its canonical form in Newton's first law of motion, which states that in the absence of external forces, a body will move in a straight line at constant velocity.  This ``law of inertia,'' as Newton called it, is preserved, \emph{mutatis mutandis}, in general relativity, where inertial motion is governed by the \emph{geodesic principle}.  The geodesic principle states that in the absence of external forces, the possible trajectories through four dimensional spacetime of a massive test point particle will be timelike geodesics---i.e., bodies will move along ``locally straightest'' lines, without acceleration.  In standard presentations of general relativity, the geodesic principle is stated as a postulate \citep[cf. ][]{Hawking+Ellis, Wald, MTW, MalamentGR}, much like Newton's first law.\footnote{For a detailed and enlightening discussion of the status of the first law of motion in standard Newtonian gravitation, see \citet{Earman+Friedman}.}  However, shortly after Einstein presented the theory, he and others began to suspect that one could equally well conceive of the geodesic principle as a theorem, at least in the presence of other standard assumptions of relativity theory \citep{Eddington, Einstein+Grommer, Einstein+etal}.  This shift from geodesic-principle-as-postulate to geodesic-principle-as-theorem has led to a widespread and deeply influential view that general relativity has a special explanatory virtue that distinguishes it from other theories of space and time: in the words of Harvey Brown, general relativity ``...is the first in the long line of dynamical theories... that \emph{explains} inertial motion'' \citep[pg. 163]{Brown}.  The view holds that Newtonian physics may answer the ``how'' part of Aristotle's question, but there is a sense in which only general relativity answers the ``why'' part.

Although Einstein's early attempts to prove the geodesic principle were not unambiguously successful, more recent efforts have shown that there is a precise sense in which the geodesic principle may be understood as a theorem of general relativity \citep{Geroch+Jang}.\footnote{There have been several steps along the way to proving the geodesic principle as a rigorous theorem of general relativity.  The most significant early attempt was the work of \citet{Einstein+Grommer} and \citet{Einstein+etal}, with subsequent work due to \citet{Taub}, \citet{Thomas}, \citet{Newman+Posadas1, Newman+Posadas2}, and \citet{Dixon}.  These are described and criticized briefly in \citet{Geroch+Jang} (see also \citet{Blanchet} and \citet{Damour} for reviews of approaches to the ``problem of motion'' in general relativity).  There are currently two approaches to the problem that are widely recognized as successful: the one developed by Geroch and Jang \citep[see also][]{Ehlers+Geroch}, which will be my focus in the present paper, and one developed by \citet{Sternberg} and \citet{Souriau}, among others, which models a massive test point particle as an order-zero distribution with support along a curve.  One can then show that if the distribution is (weakly) conserved, the curve must be a geodesic.  Although the Geroch-Jang approach and the Sternberg-Souriau approach are \emph{prima facie} different, there is a sense in which they turn out to be equivalent \citep{Geroch+Weatherall}.}     However, it turns out that relativity is not unique in this regard.  Geometrized Newtonian gravitation (sometimes, Newton-Cartan theory) is a reformulation of Newtonian gravitation due to \'Elie \citet{Cartan1, Cartan2} and Kurt \citet{Friedrichs} that shares many of the qualitative features of general relativity.  In geometrized Newtonian gravitation one represents space and time as a four dimensional spacetime manifold, the curvature of which depends dynamically on the distribution of matter on the manifold.  Gravitational influences, meanwhile, are not understood as forces, as in traditional formulations of Newtonian gravitation; rather, they are a manifestation of the curvature of spacetime.  And in particular, inertial motion is governed by the geodesic principle: in the absence of external (non-gravitational) forces, bodies move along the geodesics of (curved) spacetime.  Recently, I have shown that the geodesic principle can be understood as a theorem of geometrized Newtonian gravitation \citep{WeatherallJMP}.  Mathematically, the Newtonian theorem is nearly identical to the Geroch-Jang theorem.  Moreover, as I have argued elsewhere, when the background assumptions needed to prove these theorems are examined in the contexts of each theory, one can reasonably conclude that the geodesic principle has essentially the same status in both cases, though in neither theory is the situation as simple as one might have hoped \citep{WeatherallSHPMP}.

One consequence of this recent work is that Einstein and others' idea that the status of the geodesic principle in general relativity distinguishes the theory from other theories of space and time seems more difficult to hold on to.  But it also raises a related issue.  When one attends carefully to the details of these theorems, several complications arise concerning the strength and status of the assumptions necessary for proving them.  Given these complications, one might reasonably ask, do either of these theories \emph{explain} inertial motion?  It is this second question that I will take up in the present paper.\footnote{The recent literature on whether and in what sense general relativity and Newtonian gravitation explain inertial motion originates with Harvey \citet{Brown}.  Brown is not especially concerned to give an ``account'' of the sense of explanation he has in mind, in the sense of providing necessary or sufficient conditions for when some argument, theorem, etc. is an explanation (nor, I should say, am I!), though the idea is that the geodesic principle is explained in general relativity because there is a sense in which it is a consequence of the central dynamical principle of the theory, Einstein's equation.  Ad\'an \citet{Sus} has expanded on this view, calling the form of explanation at issue ``dynamical explanation'', and further defending Brown's claim that general relativity is distinguished from other spacetime theories with regard to the explanation it provides of inertial motion.  David \citet{MalamentGP} and I \citep{WeatherallFoP, WeatherallSHPMP}, meanwhile, have pointed out that the geodesic principle does \emph{not} follow merely from Einstein's equation, and that a strong energy condition is also required; moreover, as I note above, a theorem remarkably similar to the one that holds in the relativistic case also holds in geometrized Newtonian gravitation.  But these latter discussions largely set aside the question of what sense of explanation is at issue, if any.  More recently, Mike \citet{Tamir} has pointed out that in general relativity, at least, the geodesic principle is false for realistic matter.  He then considers almost-geodesic motion as a kind of universal phenomenon in the sense of \citet{Batterman}.  From this latter perspective, these theorems provide explanations in the sense of showing how certain behavior can be expected to arise approximately for a wide variety of substances.  The remarks in the present paper are of a rather different character than (most of) this earlier work, and so I will not engage with it closely in the text.}

I will begin with a brief overview of geometrized Newtonian gravitation, after which I will review the relevant theorems concerning the geodesic principle in that theory and general relativity.  I will focus on the subtle ways in which the theorems differ, and on the complications that arise when one tries to interpret them.  Once this background material has been laid out, I will turn to the question at hand.  The starting point for this discussion will be to observe that on one way of thinking about explanation in scientific theories, the answer to the question is ``no'': neither of these theories explains inertial motion, at least if the assumptions going into the theorems have the character I describe.  I want to resist this view, however, because I think it takes for granted that one can make clear distinctions between ``levels'' or ``tiers'' of fundamentality of the central principles of a theory.  Careful analysis of the geodesic principle theorems, meanwhile, suggests that there is another way of thinking about how the principles of a theory fit together.  The alternative view I will develop---I will call it the ``puzzle ball view'' or, perhaps more precisely, the ``puzzle ball conjecture''---holds that the foundations of physical theories, or at least \emph{these} physical theories, are best conceived as a network of mutually interdependent principles, rather than as a collection of independent and explanatorily fundamental ``axioms'' or ``postulates''.  On this view, one way to provide a satisfactory explanation of a central principle of a theory, such as the geodesic principle in general relativity or geometrized Newtonian gravitation, would be to exhibit its dependence on the other central principles of the theory, i.e., to show how the principle-to-be-explained is a consequence of the other central principles and basic assumptions of the theory.  And this is precisely what the theorems I will describe do.  And so, I will argue that there \emph{is} a sense in which both theories explain inertial motion, though some care is required to say what is meant by ``explain'' in this context.

I should be clear from the start: the language of explanation is a convenient one, but I am not ultimately interested in the semantics of the word ``explain''.  The goal is not to argue whether one thing or another is \emph{really} an explanation.  The dialectic, rather, is as follows.  Many people have suggested that general relativity provides an important kind of insight with regard to inertial motion, something to be valued and sought after in our physical theories.  One might call this thing an ``explanation'', or not.   The point, though, is that when one looks in detail at just what one gets in relativity theory (and in geometrized Newtonian gravitation), it seems to work in a different way than one might have initially guessed it would.  One response to this observation would be to say that we have not actually gotten what we were promised---or, in the language above, that general relativity does \emph{not} explain inertial motion.  But another response is to try to better understand what we \emph{do} get.  My principal thesis is that if one takes this second path, an alternative picture emerges of how the foundations of theories work.  And on this alternative picture, general relativity and geometrized Newtonian gravitation both do provide an important and very useful kind of insight into inertial motion, and more, there are clear reasons why one should value and seek out this sort of insight.  Indeed, one might even think that what we ultimately get is what we should have wanted in the first place.  I am inclined to use the word ``explanation'' for this sort of insight, but fully recognize that this usage may seem non-standard or incorrect to some readers.

\section{Overview of geometrized Newtonian gravitation}

Geometrized Newtonian gravitation is best understood as a translation of Newtonian gravitation into the language of general relativity, a way of making Newtonian physics look as much like general relativity as possible, for the purposes of addressing comparative questions about the two theories.\footnote{This brief overview of geometrized Newtonian gravitation is neither systematic or complete.  The best available treatment of the subject is given in \citet{MalamentGR}; see also \citet{Trautman}.  My notation and conventions here follow Malament's.}  The result is a theory that is strikingly similar in many qualitative respects to general relativity, but which differs in certain crucial details.  Recall that in general relativity, a \emph{relativistic spacetime} is an ordered pair $(M, g_{ab})$, where $M$ is a smooth four dimensional manifold and $g_{ab}$ is a smooth Lorentzian metric on the manifold.  In geometrized Newtonian gravitation, meanwhile, one similarly starts with a smooth four dimensional manifold $M$, but one endows this manifold with a different metric structure.  Specifically, one defines \emph{two} (degenerate) metrics.  One, a \emph{temporal metric} $t_{ab}$, has signature $(1,0,0,0)$.  It is used to assign temporal lengths to vectors on $M$: the temporal length of a vector $\xi^a$ at a point $p$ is $(t_{ab}\xi^a\xi^b)^{1/2}$.  Vectors with non-zero temporal length are called \emph{timelike}; otherwise, they are called \emph{spacelike}.  The second metric is a \emph{spatial metric} $h^{ab}$, with signature $(0,1,1,1)$.  In general one requires that these two metrics satisfy an orthogonality condition, $h^{ab}t_{ab}=\mathbf{0}$.  It is important that the temporal metric is written with covariant indices and the spatial metric with contravariant indices: since both metrics have degenerate signatures, they are not invertible, and so in general one cannot use either to raise or lower indices.  In particular, this means that the spatial metric cannot be used to assign spatial lengths to vectors directly.  Instead, one uses the following indirect method.  Given a spacelike vector $\xi^a$, one can show that there always exists a (non-unique) covector $u_a$ such that $\xi^a=h^{ab}u_b$.  One then defines the spatial length of $\xi^a$ to be $(h^{ab}u_au_b)^{1/2}$, which can be shown to be independent of the choice of $u_a$.

Given a Lorentzian metric $g_{ab}$ on a manifold $M$, there always exists a unique covariant derivative operator $\nabla$ that is compatible with $g_{ab}$ in the sense that $\nabla_ag_{bc}=\mathbf{0}$.  This does not hold for the degenerate Newtonian metrics.  Instead, there are an uncountably infinite collection of derivative operators that satisfy the compatibility conditions $\nabla_a t_{bc}=\mathbf{0}$ and $\nabla_a h^{bc}=\mathbf{0}$.  This means that to identify a model of geometrized Newtonian gravitation, one needs to specify a derivative operator in addition to the metric fields.  Thus, we define a \emph{classical spacetime} as an ordered quadruple $(M,t_{ab},h^{ab},\nabla)$, where $M$, $t_{ab}$, $h^{ab}$, and $\nabla$ are as described, the metrics satisfy the orthogonality condition, and the metrics and derivative operator satisfy the compatibility conditions.  A classical spacetime is the analog of a relativistic spacetime.  Note that the signature of $t_{ab}$ guarantees that at any point $p$, one can find a covector $t_a$ such that $t_{ab}=t_at_b$; in cases where such a field can be defined globally, we call the associated spacetime \emph{temporally orientable}.  In what follows, we will always restrict attention to temporally orientable spacetimes, and will replace $t_{ab}$ with $t_a$ whenever we specify a classical spacetime.

In both theories, timelike curves---curves whose tangent vector field is always timelike---represent the possible trajectories of point particles (and idealized observers).  And as in general relativity, matter fields in geometrized Newtonian gravitation are represented by a smooth symmetric rank-2 field $T^{ab}$ (with contravariant indices).  In general relativity, this field is called the \emph{energy-momentum tensor}; in geometrized Newtonian gravitation, it is called the \emph{mass-momentum tensor}.  The reason for the difference concerns the interpretations of the fields.  In relativity theory, the four-momentum density of a matter field with energy-momentum tensor $T^{ab}$ is only defined relative to some observer's state of motion: given an observer whose worldline has (timelike) tangent field $\xi^a$, the four-momentum density $P^a$ as determined by the observer is given by $P^a=T^{ab}\xi_b$.  When $P^a$ is timelike or null, one can define the mass density $\rho$ of the field at a point, relative to the observer, as the length of $P^a$.  Moreover, the four-momentum field can be further decomposed (relative to $\xi^a$) as $P^a=E \xi^a + p^a$, where $E=P^n\xi_n$ is the relative energy density as determined by the observer, and $p^a=P^n(\delta^a_n-\xi^a\xi_n)$ is the relative three-momentum density.  Thus, the field $T^{ab}$ encodes the relative mass, relative energy, and relative momentum densities as determined by any observer.  In geometrized Newtonian gravitation, meanwhile, \emph{all} observers make the same determination of the four-momentum density of a matter field at a point: for any observer, $P^a$ is given by $P^a=T^{ab}t_b$.  Given a particular observer whose worldline has tangent field $\xi^a$, though, one can decompose $P^a$ as $P^a=\rho\xi^a+p^a$, where $\rho=P^at_a(=T^{ab}t_a t_b)$ is the (observer-independent) mass density associated with the matter field, and where $p^a=P^n(\delta^a_n-\xi^at_n)$ is the relative three-momentum density of the matter field as determined by the observer.  Thus in geometrized Newtonian gravitation, $T^{ab}$ encodes the (absolute) mass density of a matter field, as well as its momentum relative to any observer.\footnote{Note that is general relativity, one makes a distinction between the mass and energy densities relative to a given observer, where relative mass density is the length of the four-momentum density determined by an observer at a point ($\rho=(P^aP_a)^{1/2}$) and relative energy density is $E=T^{ab}\xi_a\xi_b=P^a\xi_a$, where $\xi^a$ is the tangent field to the observer's worldline.  In geometrized Newtonian gravitation, this distinction collapses: both (absolute) mass density and (absolute) energy density are given by $\rho=T^{ab}t_at_b=P^at_b$.}

It is standard in both theories to limit attention to matter fields that satisfy several additional constraints.  In particular, in both cases one assumes that matter fields satisfy the \emph{conservation condition}, which states that their energy/mass-momentum fields are divergence free (i.e., $\nabla_aT^{ab}=\mathbf{0}$).  One also usually requires that such fields satisfy various \emph{energy conditions}.  In geometrized Newtonian gravitation, only one such condition is standard: it is the so-called \emph{mass condition}.
\begin{quote}\singlespacing\textbf{Mass condition}: A mass-momentum field satisfies the mass condition if, at any point, either $T^{ab}=\mathbf{0}$ or $T^{ab}t_at_b>0$.\end{quote}  Since $T^{ab}t_at_b=\rho$ is the mass density, this assumption states that whenever the mass-momentum tensor is non-vanishing, the associated matter field has positive mass.  The situation is more complicated in general relativity, where there are several energy conditions that one may consider.  I will mention a few because they are of particular interest for present purposes.  One, called the \emph{weak energy condition}, is (at least \emph{prima facie}) quite similar to the mass condition it states that the energy density of a matter field as determined by any observer is always non-negative.
\begin{quote}\singlespacing\textbf{Weak energy condition}: An energy-momentum field satisfies the weak energy condition if, give any timelike vector $\xi^a$ at a point, $T^{ab}\xi_a\xi_b\geq 0$.\end{quote}
It is also common to consider stronger conditions.  For instance, there are the \emph{dominant energy condition} and the \emph{strengthened dominant energy condition}:
\begin{quote}\singlespacing\textbf{Dominant Energy Condition}: An energy-momentum field satisfies the dominant energy condition if, given any timelike vector $\xi_a$ at a point, $T^{ab}\xi_a\xi_b\geq0$ and $T^{ab}\xi_a$ is timelike or null.\end{quote}
\begin{quote}\singlespacing\textbf{Strengthened Dominant Energy Condition}: An energy-momentum field satisfies the strengthened dominant energy condition if, give any timelike covector $\xi_a$ at any point in $M$, $T^{ab}\xi_a\xi_b\geq 0 $ and either $T^{ab}=\mathbf{0}$ or $T^{ab}\xi_a$ is timelike.
\end{quote}
If these two conditions obtain for some matter field, then not only do all observers take the field to have non-negative energy density, they also take its four-momentum to be causal or timelike (respectively).  In other words, these latter conditions capture the requirement that matter must propagate at or below the speed of light.

The curvature of a classical spacetime is defined in the standard way: given a derivative operator $\nabla$, the \emph{Riemann curvature tensor} $R^a{}_{bcd}$ is the unique tensor field such that for any vector field $\xi^a$, $R^a{}_{bcd}\xi^b=-2\nabla_{[c}\nabla_{d]}\xi^a$.  The \emph{Ricci curvature tensor}, meanwhile, is given by $R_{ab}=R^n{}_{abn}$.  In both contexts, one says that a spacetime is \emph{flat} if $R^a{}_{bcd}=\mathbf{0}$; in geometrized Newtonian gravitation, one also says that a (possibly curved) spacetime is \emph{spatially flat} if $R^{abcd}=R^a{}_{mno}h^{bm}h^{cn}h^{do}=\mathbf{0}$ or, equivalently, $R_{mn}h^{ma}h^{nb}=\mathbf{0}$.   Given these ingredients, one can state the sense in which in geometrized Newtonian gravitation, the curvature of spacetime depends on the distribution of matter: namely, the central dynamical principle of the theory, the \emph{geometrized Poisson equation}, states that $R_{ab}=4\pi\rho t_at_b$, where $\rho$ is the mass-density defined above.  This expression explicitly relates the Ricci curvature of spacetime to the distribution of matter.  It is the Newtonian analogue of Einstein's equation, $R_{ab}=8\pi(T_{ab}-\frac{1}{2}Tg_{ab})$, where $T=T^{ab}g_{ab}$, or equivalently $8\pi T_{ab}=R_{ab}-\frac{1}{2}Rg_{ab}$, where $R=R_{ab}g^{ab}$.

There are a few points to emphasize here concerning the geometrized Poisson equation.  For one, if the geometrized Poisson equation holds of a classical spacetime for some mass-momentum tensor $T^{ab}$, then the classical spacetime is spatially flat, since $R_{nm}h^{na}h^{mb}=4\pi\rho t_nt_mh^{ma}h^{nb}=\mathbf{0}$.  This fact is a way of recovering a familiar feature of Newtonian gravitation, namely that \emph{space} is always flat, even though in the geometrized theory \emph{spacetime} may be curved.   Second, in general relativity one can freely think of both the metric and the derivative operator as (systemically related) dynamical variables in the theory.  In geometrized Newtonian gravitation, this is not the case: instead, the metrical structure of a classical spacetime is fixed, and only the derivative operator (or more specifically, the Ricci curvature, which is defined in terms of the derivative operator) is a dynamic variable.  Finally, there is a sense in which, given some matter distribution, the geometrized Poisson equation ``fixes'' a derivative operator on a classical spacetime, but one has to be careful, as one can typically only recover a unique derivative operator satisfying the geometrized Possion equation for a given matter distribution in the presence of additional boundary conditions or other assumptions.

The geometrized Poisson equation provides the sense in which in geometrized Newtonian gravitation, spacetime is curved in the presence of matter; the sense in which gravitational effects may be understood as a manifestation of this curvature is just the same as in general relativity.  That is, a derivative operator allows one to define a class of geometrically privileged curves, the \emph{geodesics} of the spacetime, which consist of all curves whose tangent fields $\xi^a$ satisfy $\xi^n\nabla_n\xi^a=\mathbf{0}$ everywhere.  I have already said that the timelike curves of a spacetime represent the possible trajectories for massive particles; the timelike geodesics, meanwhile, represent the possible \emph{unaccelerated} trajectories of particles in both theories.  The geodesic principle then connects these geometrically privileged curves with force-free motion.  Thus, in geometrized Newtonian gravitation, as in general relativity, the distribution of matter throughout space and time affects the possible trajectories of massive point particles not by causing such particles to accelerate, but rather by dynamically determining a collection of unaccelerated curves.

These features of geometrized Newtonian gravitation provide the sense in which the theory is qualitatively similar to general relativity.  But one might wonder what undergirds the implicit claim that geometrized Newtonian gravitation is in some sense \emph{Newtonian}.  One sense in which the theory is Newtonian is immediate: the degenerate metric structure of a classical spacetime captures the implicit geometry of space and time in ordinary Newtonian gravitation, where one has a temporally ordered succession of flat three dimensional manifolds representing space at various times \citep[cf. ][]{SteinNST}.  But there is more to say.  In standard formulations of Newtonian gravitation, spacetime is flat.  Gravitation is a force mediated by a gravitational potential, which in turn is related to the distribution of matter by Poisson's equation.  In the present four dimensional geometrical language, this can be expressed as follows.  We begin with a classical spacetime $(M,t_a,h^{ab},\nabla)$ as before, but now we require that $\nabla$ is flat, i.e., $R^a{}_{bcd}=\mathbf{0}$.  We again represent matter by its mass-momentum field $T^{ab}$, defined just as above, but we also define a scalar field $\varphi$, which is the gravitational potential.  Poisson's equation is written as $\nabla^a\nabla_a\varphi=4\pi\rho$ where the index on $\nabla^a$ is raised using $h^{ab}$, and where $\rho=T^{ab}t_at_b$.  And now the acceleration of a massive test point particle in the presence of a gravitational potential $\varphi$ is given by $\xi^n\nabla_n\xi^a=-\nabla^a\varphi$, where $\xi^a$ is the tangent to the particle's trajectory.  In other words, in standard Newtonian gravitation matter accelerates in the presence of mass.

\begin{figure}[h]
\centering
\includegraphics[width=.85\columnwidth]{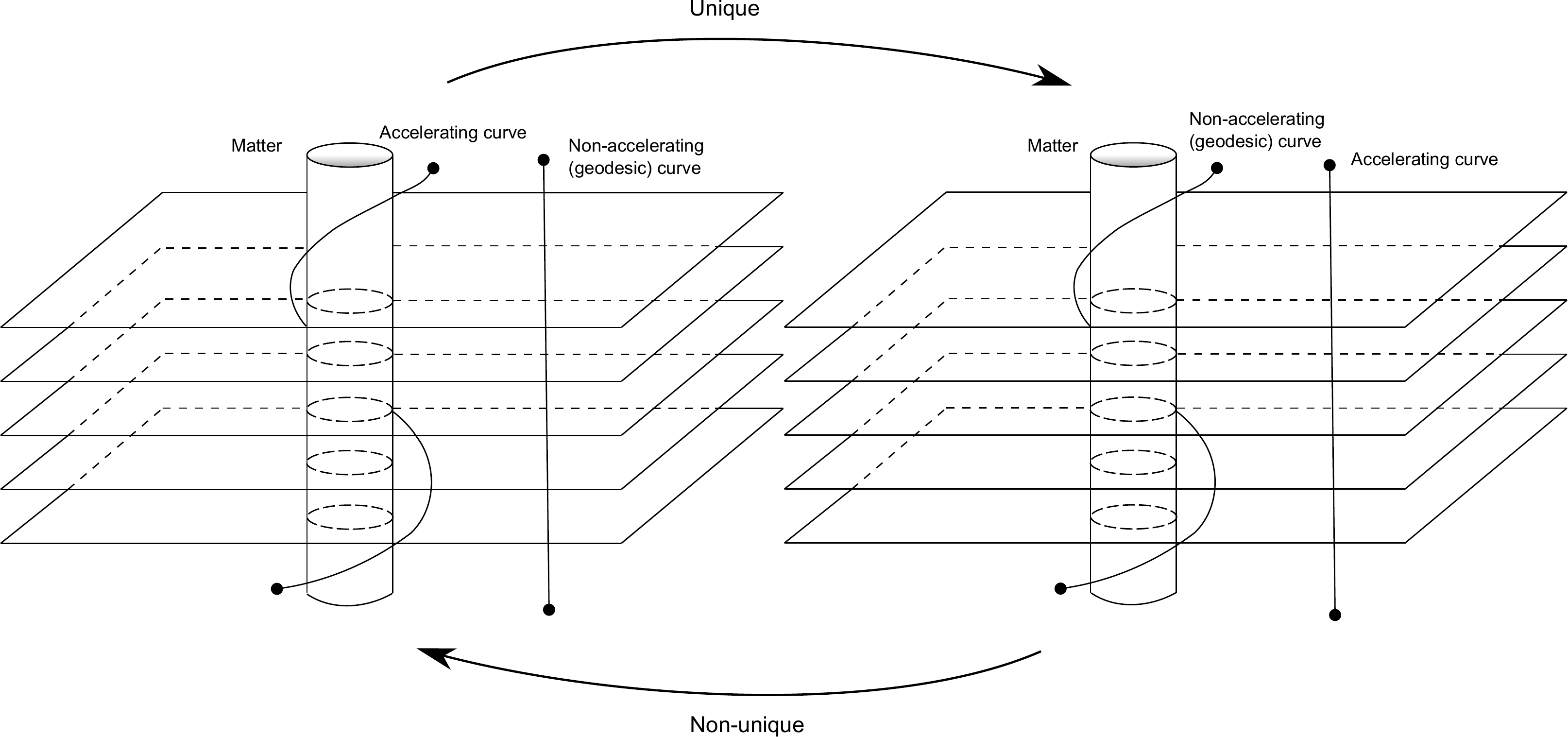}
\caption{\label{translate} In general it is possible to translate between geometrized Newtonian gravitation and standard Newtonian gravitation, as depicted in this figure.  On the left is a model of standard Newtonian gravitation: one has a matter field represented by the world-tube of some body, such as the sun, and a curve orbiting this body, representing, say, a small planet.  This curve corresponds to an allowed trajectory insofar as it is accelerating by the appropriate amount. On the right is the corresponding model of geometrized Newtonian gravitation.  One has precisely the same matter distribution, and the same allowed trajectory (i.e., the same orbit), but now we understand this trajectory to be allowed by the theory because it is a geodesic of a curved derivative operator, with curvature determined by the matter distribution.  Note that both theories have the same metrical structure, represented here by a succession of flat slices representing space at various times.}
\end{figure}

It turns out that standard Newtonian gravitation (thus understood) and geometrized Newtonian gravitation are systematically related \citep[ch. 4.2]{MalamentGR}.  Specifically, given a classical spacetime $(M,t_a,h^{ab},\nabla)$ with $\nabla$ flat, a smooth mass density $\rho$, and a smooth gravitational potential $\varphi$ satisfying $\nabla^a\nabla_a\varphi=4\pi\rho$, there always exists a unique derivative operator $\tilde{\nabla}$ such that $(M,t_a,h^{ab},\tilde{\nabla})$ is a classical spacetime, $\tilde{R}_{ab}=4\pi\rho t_a t_b$, and such that for any timelike vector field $\xi^a$, $\xi^n\nabla_n\xi^a=-\nabla^a\varphi$ if and only if $\xi^n\tilde{\nabla}_n\xi^a=\mathbf{0}$.  In other words, given a model of standard Newtonian gravitation, there is always a model of geometrized Newtonian gravitation with precisely the same mass density and allowed trajectories.  Additionally, the derivative operator $\tilde{\nabla}$ will always satisfy two curvature conditions: $\tilde{R}^{ab}{}_{cd}=\mathbf{0}$ and $\tilde{R}^a{}_b{}^c{}_d=\tilde{R}^c{}_d{}^a{}_b$.  This result is known as the Trautman geometrization lemma; it provides the sense in which one can always translate from standard Newtonian gravitation into the geometrized theory.  One can also prove a corresponding recovery lemma (also due to Trautman), allowing for translations back: namely, given a classical spacetime $(M,t_a,h^{ab},\tilde{\nabla})$ and smooth mass density $\rho$ satisfying $\tilde{R}_{ab}=4\pi\rho t_a t_b$, if $\tilde{R}^{ab}{}_{cd}=\mathbf{0}$ and $\tilde{R}^a{}_b{}^c{}_d=\tilde{R}^c{}_d{}^a{}_b$ then there always exists a flat derivative operator $\nabla$ and a gravitational potential $\varphi$ such that $(M,t_a,h^{ab},\nabla)$ is a classical spacetime, $\nabla^a\nabla_a\varphi=4\pi\rho$, and again for any timelike vector field $\xi^a$, $\xi^n\nabla_n\xi^a=-\nabla^a\varphi$ if and only if $\xi^n\tilde{\nabla}_n\xi^a=\mathbf{0}$.  Note that this recovery result only holds in the presence of the two additional curvature conditions stated above; moreover, in general the translation from geometrized Newtonian gravitation to standard Newtonian gravitation will \emph{not} be unique.  (See figure \ref{translate}.)

\section{The geodesic principle as a theorem}

With the background of the previous section in place, I can now state the precise sense in which the geodesic principle may be understood as a theorem in general relativity and geometrized Newtonian gravitation.  I will begin by stating both theorems, and then double back to the question of how one should interpret them.
\begin{thm}\label{GJ} \emph{\textbf{\citep{Geroch+Jang}}}\footnote{This particular statement of the theorem is heavily indebted to \citet[Prop. 2.5.2]{MalamentGR}.}Let $(M,g_{ab})$ be a relativistic spacetime, and suppose $M$ is oriented. Let $\gamma:I\rightarrow M$ be a smooth imbedded curve.  Suppose that given any open subset $O$ of $M$ containing $\gamma[I]$, there exists a smooth symmetric field $T^{ab}$ with the following properties.
\begin{enumerate}
\item \label{sdec} $T^{ab}$ satisfies the \emph{strengthened dominant energy condition}, i.e., given any timelike covector $\xi_a$ at any point in $M$, $T^{ab}\xi_a\xi_b\geq 0$ and either $T^{ab}=\mathbf{0}$ or $T^{ab}\xi_a$ is timelike;
\item \label{cons}$T^{ab}$ satisfies the \emph{conservation condition}, i.e., $\nabla_a T^{ab}=\mathbf{0}$;
\item \label{inside}$\supp{T^{ab}}\subset O$; and
\item \label{non-vanishing}there is at least one point in $O$ at which $T^{ab}\neq \mathbf{0}$.
\end{enumerate}
Then $\gamma$ is a timelike curve that can be reparametrized as a geodesic.
\end{thm}
One can prove an almost identical theorem in geometrized Newtonian gravitation.
\begin{thm} \emph{\textbf{ \citep{WeatherallJMP}}}
\label{W}
Let $(M,t_{ab},h^{ab},\nabla)$ be a classical spacetime, and suppose that $M$ is oriented.  Suppose also that $R^{ab}{}_{cd}=\mathbf{0}$.  Let $\gamma:I\rightarrow M$ be a smooth imbedded curve.  Suppose that given any open subset $O$ of $M$ containing $\gamma[I]$, there exists a smooth symmetric field $T^{ab}$ with the following properties.
\begin{enumerate}
\item\label{mass} $T^{ab}$ satisfies the mass condition, i.e., whenever $T^{ab}\neq \mathbf{0}$, $T^{ab}t_at_b>0$;
\item\label{cons2} $T^{ab}$ satisfies the conservation condition, i.e., $\nabla_a T^{ab}=\mathbf{0}$;
\item\label{inside2} $\supp{T^{ab}}\subset O$; and
\item\label{non-vanishing2} there is at least one point in $O$ at which $T^{ab}\neq \mathbf{0}$.
\end{enumerate}
Then $\gamma$ is a timelike curve that can be reparametrized as a geodesic.
\end{thm}

As a first remark, it may not be obvious that either of these theorems should be understood to capture the geodesic principle at all, at least in a natural way.    A principal difficulty in trying to derive the geodesic principle as a theorem concerns a kind of ontological mismatch between the geodesic principle and the rest of general relativity: namely, general relativity is a field theory, whereas the geodesic principle is a statement concerning point particles.  One strategy for dealing with this problem is to try to model massive point particles as ``small'' bits of extended matter, and then try to show that under sufficiently general assumptions, the world tubes of such small bits of matter will contain timelike geodesics.  But this turns out to be false in general---geodesic motion only obtains in the idealized limit where the worldtube of a body collapses to a curve, in which case one can no longer represent matter as a smooth field on spacetime.\footnote{This point is emphasized by \citet{Tamir}.}  The Geroch-Jang strategy, meanwhile, is different.  Instead of starting with some matter and asking what kind of trajectory it follows, one starts with a curve and asks under what circumstances that curve can be understood as a trajectory for arbitrarily small bits of extended matter.  Both theorems then state that the only curves along which arbitrarily small bits of matter can be constructed are timelike geodesics.

Importantly, one represents a ``small bit of matter'' by a smooth symmetric rank 2 tensor fields with support in some neighborhood of the curve.  But a curve is not understood as a possible trajectory for a free massive test point particle if one can construct \emph{any} smooth symmetric rank 2 tensor field in arbitrarily small neighborhoods of the curve---rather, one limits attention to fields that satisfy additional constraints.  The claim, then, is that these theorems capture the geodesic principle in both theories insofar as the additional constraints on the matter fields adequately capture what we intend by ``free massive test matter''. This means that the interpretation of the theorems turns on the status of these conditions.  And so, for a comparative study of the status of the geodesic principle in each theory, one wants to compare the status of each of these assumptions relative to their respective theories.

Two of the assumptions can be set aside immediately: in both theorems, assumptions (3) and (4) play the role of setting up the limiting process implicit in the theorems.  Assumption (3) limits attention to matter fields that vanish outside one's chosen neighborhood of the curve (which captures the sense in which one is considering arbitrarily small bits of matter propagating along the curve), and assumption (4) indicates that the matter field must be non-vanishing somewhere along curve, ruling out the trivial case.  These assumptions are identical in both cases, and neither is troublingly strong.  There is also an obvious difference that can be safely ignored.  In the Newtonian theorem, we place an additional constraint on the curvature, namely $R^{ab}{}_{cd}=\mathbf{0}$.  This is precisely the curvature condition needed to prove the Trautman recovery theorem, allowing one to translate from a model of geometrized Newtonian gravitation to a model of standard Newtonian gravitation.  For this reason, the curvature condition is naturally interpreted as a restriction to models of geometrized Newtonian gravitation that are \emph{Newtonian}, in the sense that they admit translations back to models of standard Newtonian gravitation.  This presumably is the case of greatest interest, and so I am inclined to think of the assumption as benign.\footnote{For another view on this matter, see \citet{Sus}.}  Moreover, there is good reason to think that this assumption can be dropped, though to my knowledge, proving as much is still an open (and perhaps interesting) problem.

The most striking difference between the two theorems concerns the respective assumptions (1).  In the Newtonian theorem, this is the mass condition, i.e., that whenever the mass-momentum field is non-vanishing, the mass density determined by any observer must be positive.  This is the standard energy condition in geometrized Newtonian gravitation, and more, it is natural in this context, as it captures the sense in which the bits of matter being represented are massive.  In the Geroch-Jang theorem, meanwhile, one requires the strengthened dominant energy condition, which states that (a) all observers must assign non-negative (mass-)energy density to the matter field (the weak energy condition) and (b) that if $T^{ab}\neq\mathbf{0}$, then the four-momentum assigned to the matter field by any observer must by timelike.  It seems natural to think that the \emph{weak} energy condition, (a), is playing the role played by the mass condition in the Newtonian case: namely, it captures the sense in which the small bits of matter are massive, by requiring that they always have non-negative mass.  But from this perspective, the second part of the condition, (b), is a strong additional requirement.  In Newtonian gravitation, it would seem, one needs only to assume that mass is always positive to get timelike geodesic propagation, whereas in general relativity, one also needs to make an assumption about the timelike propagation of matter.\footnote{This is precisely how I present the situation in \citet{WeatherallSHPMP} and \citet{WeatherallFoP}.  However, I now think matters are still more complicated than I indicate there, as I explain in the text.  Still, the principal morals of those previous discussions are unchanged by these additional considerations.}

However, the situation is not quite so simple as this. Although the mass condition appears to be nothing more than an assumption about positive mass, it, too, contains an implicit assumption about timelike propagation.  To see this, consider a different (non-standard) Newtonian energy condition, which I will call the \emph{weakened mass condition}.
\begin{quote}\singlespacing\textbf{Weakened Mass Condition}: A mass-momentum field $T^{ab}$ satisfies the weakened mass condition if at any point, $T^{ab}t_at_b\geq\mathbf{0}$.\end{quote}
The weakened mass condition has a good claim on being the Newtonian analogue of the weak energy condition and might similarly be understood as the claim that mass/energy density is always non-negative.   But it is \emph{strictly weaker} than the mass condition, since the weakened mass condition may be satisfied by mass-momentum fields that are spacelike, in the sense that $T^{ab}\neq\mathbf{0}$ but $T^{ab}t_at_b=0$ (for example, consider $T^{ab}=u^au^b$, with $u^a$ a spacelike vector field).  In other words, the mass condition amounts to the weakened mass condition plus the additional assumption that $T^{ab}$ is timelike.  We can make this explicit by defining an equivalent condition, the \emph{modified mass condition}.
\begin{quote}\singlespacing\textbf{Modified Mass Condition}: A mass-momentum field $T^{ab}$ satisfies the modified mass condition if at any point, $T^{ab}t_at_b\geq\mathbf{0}$ and either $T^{ab}=\mathbf{0}$ or $T^{ab}t_a$ is timelike.\end{quote}
The modified mass condition is equivalent to the mass condition, but would appear to be the natural translation of the strengthened dominant energy condition.  On the other hand, one can also rewrite the weak energy condition as the \emph{strengthened weak energy condition}.
\begin{quote}\singlespacing\textbf{Strengthened Weak Energy Condition}: An energy-momentum field satisfies the strengthened weak energy condition if, give any timelike vector $\xi^a$ at a point, either $T^{ab}=\mathbf{0}$ or $T^{ab}\xi_a\xi_b> 0$.\end{quote}
 stating that for all timelike $\xi^a$, if $T^{ab}\neq\mathbf{0}$, then $T^{ab}\xi_a\xi_b>0$.  And \emph{this} condition seems like the natural translation of the (standard) mass condition, but it is strictly weaker than the strengthened dominant energy condition!\footnote{The strengthened weak energy condition is also strictly weaker than the (strict) dominant energy condition, and so Prop. 4 of \citet{WeatherallFoP} implies that the strengthened dominant energy condition is not strong enough to prove the Geroch-Jang theorem.}

This situation is summarized in table \ref{energyConditions}.
\begin{center}
\begin{table}[h]
\begin{tabular}{|c c c|}
\hline
Geometrized Newtonian gravitation & &General relativity\\
\hline
Modified mass condition & $\longleftrightarrow$ & Strengthened dom. energy condition\\
$\Updownarrow$ & & $\Downarrow$\\
Mass condition & $\longleftrightarrow$ & Strengthened weak energy condition\\
$\Downarrow$ & & $\Downarrow$\\
Weakened mass condition & $\longleftrightarrow$ & Weak energy condition\\
\hline
\end{tabular}
\caption[The relationships between several energy conditions]{\label{energyConditions} This table summarizes the relationship between the various energy condition discussed in the text. Single arrows represent ``apparently natural translations''; double arrows represent logical implications.}
\end{table}
\end{center}
There are thus two ways of thinking about the relationship between the energy conditions used in these theorems, depending on which ``natural translations'' one emphasizes.  On one way of thinking, the mass condition is essentially the same as the strengthened weak energy condition.  From this point of view, then, the strengthened dominant energy condition in the Geroch-Jang theorem is a strictly stronger assumption than the corresponding assumption in Theorem \ref{W}.  More, one might be inclined to think that one gets something additional for free in the Newtonian case, since the mass condition turns out to imply the (apparently) stronger modified mass condition, whereas the strengthened weak energy condition does not imply the strengthened dominant energy condition.  Meanwhile, on the other way of thinking about things, one argues that the strengthened dominant energy condition is essentially the same as the modified mass condition, which is fully equivalent to the mass condition.  And so one concludes that the energy conditions required by the two theorems are essentially the same.

There is also another possibility, which is to say that one cannot perform simple translations between the energy conditions in these two theories at all.  I am inclined to endorse this last option, though this raises new questions about how one should compare the theorems.  There are a few things to say.  First, irrespective of how one tries (or does not try) to translate these conditions, there are still two senses in which the strengthened dominant energy condition is stronger than the mass condition.  One is that the timelike propagation clause of the strengthened dominant energy condition can be understood as the assumption that the instantaneous speed of matter, relative to any observer, must be strictly less than the speed of light.  The corresponding clause of the (modified) mass condition, meanwhile, amounts to the assumption that matter cannot propagate at \emph{infinite} speed relative to any observer.  And the assumption that a number must be less than a fixed finite value is stronger than the assumption that it must be finite, but not bounded.

The second, more significant sense in which the strengthened dominant energy condition is stronger is that the \emph{only} way in which matter in Newtonian gravitation can be ``massive'' (i.e., have positive mass as determined by some observer) is if it satisfies the mass condition.  Matter that satisfies the weakened mass condition but not the mass condition will necessarily have zero mass. And so one might argue that the mass condition is necessary to capture what is meant by ``massive'' in the context of Newtonian gravitation.  In general relatively, meanwhile, matter can be ``massive'' in two senses, \emph{without} satisfying the strengthened dominant energy condition: it can be massive in the sense that it has positive (mass-)energy density (i.e., it satisfies the weak energy condition), and it can be massive in the sense that some observers will assign it positive mass density (i.e., the relative four-momentum density as determined by some observers is timelike).  This second sense trades on an important distinction between \emph{some} observers assigning positive mass density and \emph{all} observers assigning positive mass density.  One might have thought that in order for a matter field to be massive, it would be sufficient if some observers, perhaps privileged---say co-moving observers making determinations of ``rest mass density'', when that makes sense---and perhaps not, determine that the field has positive mass density.  But the strengthened dominant energy condition requires considerably more than this.  In geometrized Newtonian gravitation, meanwhile, all of these distinction collapse.  If anyone determines a matter field has positive mass, then everyone does.

A final remark is that, understood within the context of the respective theories, the strengthened dominant energy condition is a more surprising assumption to have to make than the mass condition.  One often thinks of relativity theory as \emph{forbidding} superluminal propagation of matter, in the sense that somehow the geometric structure of the theory renders superluminal matter incoherent.  But here, at least, it seems that we need to rule out superluminal propagation of matter as an additional assumption in order to derive the geodesic principle.  This point can be made precise by asking whether one can drop or weaken the energy condition in the Geroch-Jang theorem and still derive the geodesic principle.  And the answer is ``no''. If one drops the energy condition altogether, it is possible to construct bits of matter that propagate along \emph{any} timelike curve \citep{MalamentGP}.  And if one weakens the energy condition to the weak energy condition or the dominant energy condition, one can construct bits of matter that propagate along spacelike or null curves, respectively \citep{WeatherallFoP}.  To be sure, in the Newtonian case the mass condition is similarly necessary (the considerations offered in \citet{WeatherallFoP} can be adapted to show that the weakened mass condition is not enough to get timelike geodesic motion in geometrized Newtonian gravitation), but this does not seem as striking, since one does not expect Newtonian gravitation to imply restrictions on the propagation of matter, even if it is standard to assume that matter cannot propagate instantaneously in the theory.

This leaves the conservation condition, assumption (2) in both theorems.  The statement of the assumption is identical in both cases, namely that the tensor fields representing matter must divergence free.  And in both theories, this assumption is a way of capturing that the bits of matter must be \emph{free} in the sense of non-interacting.  This interpretation is justified because in both theories there is a standard background assumption that at every point of spacetime, \emph{total} energy/mass-momentum must be divergence free, and more, that a particular energy/mass-momentum field fails to be divergence free at a point just in case it is interacting with some other such field at that point.  And so, to say that a particular field satisfies the conservation condition everywhere is to say that that field cannot be exchanging energy/mass-momentum with any other fields.

So far, it would seem that these assumptions have precisely the same status in both theorems.  But this is too quick.  Although the assumptions are equally natural ways of capturing the desired sense of ``free'' in both theorems, they only have that interpretation in the presence of the background assumption regarding the local conservation of total energy/mass-momentum.  And there is an argument to be made that this background assumption has a different status in general relativity than in geometrized Newtonian gravitation.  In general relativity, Einstein's equation implies the conservation condition, at least for total source matter.  This is because the equation can be written as $8\pi T^{ab}=R^{ab}-\frac{1}{2}g^{ab}R$, and it is a brute geometrical fact (known as Bianchi's identity) that the right-hand side of this equation is always divergence free.  Thus, the left-hand side must also be divergence free.  The geometrized Poisson equation, however, does \emph{not} imply the conservation condition.  And so, if one has Einstein's equation lurking in the background, one might be inclined to say that the background assumption that matter is conserved comes for free in general relativity, whereas it is an additional brute assumption in geometrized Newtonian gravitation.  There is an important caveat here---the argument that Einstein's equation implies the conservation condition only applies for source matter, whereas the geodesic principle is supposed to govern \emph{test} matter, i.e., matter that is not treated as a source in Einstein's equation---but nonetheless, one might think that the conservation condition has a special status---even, to anticipate the discussion in the next section, a privileged explanatory status---in general relativity because of its relation to Einstein's equation.\footnote{I should emphasize: one does not \emph{need} to think of the conservation condition as having a different status in general relativity than in geometrized Newtonian gravitation.  For instance, I have elsewhere argued that one can think of the conservation condition as a meta-principle, in the sense that the assumption that matter is conserved is expected to hold true in a wide variety of theories, and that from this perspective the status of the assumption is much the same in both general relativity and geometrized Newtonian gravitation \citep{WeatherallSHPMP}.  (Of course, the assumption that a matter field is divergence free is not exactly the same as the assumption that total mass or energy is constant over time, but it does deserve to be called the relativistic version of traditional conservation principles.)}

In the next section, I will turn to the question of whether either of these theorems should count as explanations of inertial motion.  But before I do so, it will be helpful to sum up the discussion in the present section.  I have now made precise the sense in which one can prove the geodesic principle as a theorem of both general relativity and geometrized Newtonian gravitation.  But, as I hope has become clear, interpreting and comparing these theorems is quite subtle.  It is not quite right to say that the theorems have the same interpretation or significance: on the one hand, there is arguably a sense in which the conservation condition, necessary for both theorems, has a different and perhaps privileged status in general relativity; and on the other hand, there are several senses in which the energy condition required for Geroch-Jang theorem is stronger than the condition required for the Newtonian theorem, both in absolute terms and relative to the respective theories.  Despite these differences, however, there is at least one important sense in which the status of the geodesic principle is strikingly similar in both theories.  In both cases, one \emph{can} prove the geodesic principle as a theorem.  But to do so, one needs to make strong assumptions about the nature of matter.  The status of these assumption will play a central role in what follows.

\section{Explaining inertial motion?}

General relativity and geometrized Newtonian gravitation, like any physical theory, involve a number of basic assumptions and central principles.  For instance, general relativity begins with some background assumptions about matter and geometry: space and time are represented by a four dimensional, possibly curved Lorentzian manifold; matter is represented by its energy-momentum tensor, a smooth symmetric rank two field on spacetime.  One then adds some additional assumptions, as principles indicating how to interpret and use the theory.  One may stipulate that total energy-momentum at a point must satisfy the conservation condition.  One assumes that matter fields satisfy various possible energy conditions, that idealized clocks measure proper time along their trajectories, and that free massive test point particles traverse timelike geodesics.  We postulate a dynamical relationship between the geometrical structure of spacetime and the energy-momentum field.  And so on.  Some of these assumptions involve stipulating kinematical structure; others involve basic constraints and dynamical relationships; others still tell us how to extract empirical content from the theory.  All of them have some claim to centrality or fundamentality in the theory.

But they are not necessarily independent.  For instance, as I mention above, the conservation condition may be understood as a consequence of Einstein's equation, at least for source matter.  And so, at least in some contexts, one might want to think of the conservation condition as somehow subordinate to Einstein's equation.  One might even be inclined to say that it is Einstein's equation that really deserves to be called the ``fundamental principle,'' while the conservation condition has some other, less fundamental status---or, in other words, that Einstein's equation \emph{explains} why matter is locally conserved.  We might even say that this is what it \emph{means} to say that something like the conservations condition is explained by a theory: it can be derived from  more fundamental principles in the theory.

From this point of view, one might have thought that when Einstein, Eddington, and others have claimed that general relativity explains inertial motion, in the sense that one can prove the geodesic principle as a theorem, the claim would have been analogous to what I have just said about the conservation condition: namely, one can take some collection of other principles of the theory and use them to derive the geodesic principle.  One might then think that the geodesic principle has the same subordinate status as the conservation condition.  It may be central to the theory, but not truly fundamental.  The fundamental principles are the ones that go into proving the geodesic principle.  On this view, one thinks of the foundations of general relativity as a two-tiered system.  On the top tier are the truly fundamental principles; on the lower tier are the other central principles that can be derived from the top-tier principles.  Initially, perhaps, one thought that the geodesic principle and conservation condition were top-tier principles; but the Geroch-Jang theorem and Bianchi's identity show that they are really second-tier principles.\footnote{Indeed, it seems Einstein originally \emph{did} think of the conservation condition as a top-tier principle, in the sense that he thought it was an independent assumption that any realistic field equation would need to be compatible with.  See \citet{Earman+Glymour1, Earman+Glymour2}.}

Thinking this way can lead to problems, however.  The main moral of the last section was that although one can prove the geodesic principle as a theorem in both general relativity and geometrized Newtonian gravitation, to do so one requires strong assumptions about the nature of matter.  And so, if we want to move the geodesic principle to the lower tier, it would seem that we need to understand these assumptions as as top-tier principles.  But this raises a question: why should we think of \emph{these} principles as the truly fundamental ones?  Or more specifically, why should we think of the conservation condition and the respective energy conditions as more fundamental than the geodesic principle itself?

If one were committed to the idea that the geodesic principle is a second-tier principle in one or both of these theories, perhaps one would be willing to include the assumptions needed to prove the geodesic principle among the truly fundamental principles of that theory.  But it is hard to see how this is an appealing move on independent grounds.  Even if one were to argue that dynamical principles such as Einstein's equation and the geometrized Poisson equation are clearly more fundamental than the geodesic principle, it remains the case that the strong energy condition needed to prove the Geroch-Jang theorem is entirely independent of Einstein's equation.  (And neither assumption follows from the geometrized Poisson equation.)  More, there is a sense in which one can draw all of the inferential arrows in the opposite direction, at least in one important case.  Consider an energy/mass-momentum field of the form $T^{ab}=\rho\xi^a\xi^b$, for some smooth scalar field $\rho$ and smooth vector field $\xi^a$.  An energy/mass-momentum field of this form is the natural way of representing a matter field composed of mutually non-interacting massive point particles (at least when $\rho$ is non-negative).  And so, since the geodesic principle governs the behavior of free massive test point particles, we can use it to derive features of this matter field: specifically, the geodesic principle implies that the flow-lines of the field, which represent the trajectories of each speck of dust, must be timelike geodesics.  These flow lines are just the integral curves of $\xi^a$, and so it follows that $\xi^a$ must be timelike and geodesic (i.e., $\xi^n\nabla_n\xi^a=\mathbf{0}$).  But if $\xi^a$ is timelike, then $T^{ab}$ satisfies the strengthened dominant energy condition (or respectively, the mass condition in geometrized Newtonian gravitation).  And if it is geodesic, then $T^{ab}$ is divergence free. Thus the geodesic principle allows us to derive that matter fields consisting of non-interacting massive test point particles satisfy precisely the two conditions we need to assume in order to prove the geodesic principle.\footnote{Of course, the present argument does not imply that the conservation condition and energy conditions hold for \emph{all} matter---just for the type of matter directly governed by the geodesic principle.}

So perhaps we should not be so quick to declare the conservation condition and energy conditions top-tier in either theory.  At very least, it is not perfectly clear that these assumptions are more fundamental than the geodesic principle.  But thinking in this way might lead one to conclude that \emph{neither} of the geodesic principle theorems has much explanatory significance, since (the intuition might go) explanations always proceed from more fundamental or basic facts to less fundamental facts.  Here, meanwhile, the arrows of fundamentality are muddled.  And this would mean that not only is general relativity not special with regard to its explanation of inertial motion---it does not explain inertial motion at all!

\section{The puzzle ball conjecture}

I do not find the argument I offer in the previous section compelling.  It rests on a basic intuition: to explain something like the geodesic principle, one must begin with some truly fundamental principles and then provide an argument for why the principle-to-be-explained must follow from these more fundamental ones.  This intuition takes for granted that we can make sense of a distinction between different tiers of fundamentality among the central principles of a theory like general relativity or geometrized Newtonian gravitation. And I think that this is a mistake---or at least, that there is a more compelling way of thinking about things.

Consider what the geodesic principle theorems \emph{do} accomplish.  In both of these cases, the theorems show how in the presence of other basic assumptions of the respective theories, the geodesic principle follows.  Or in other words, they show that given that one is committed to the rest of (say) general relativity, one must also be committed to the geodesic principle. One cannot freely change the geodesic principle without also changing the rest of general relaitivty: one cannot ``fiddle'' with the theory by (merely) replacing the geodesic principle with the assumption that free massive test point particles traverse some other class of curves---uniformly accelerating curves, say, or spacelike curves.  The geodesic principle is not modular, in the sense that one cannot construct a collection of perfectly good theories that differ only with how they treat inertial motion.  More, the theorems clarify precisely how it is that the geodesic principle ``fits in'' among the other central principles of general relativity.

\begin{figure}
\centering
\includegraphics[width=.4\columnwidth]{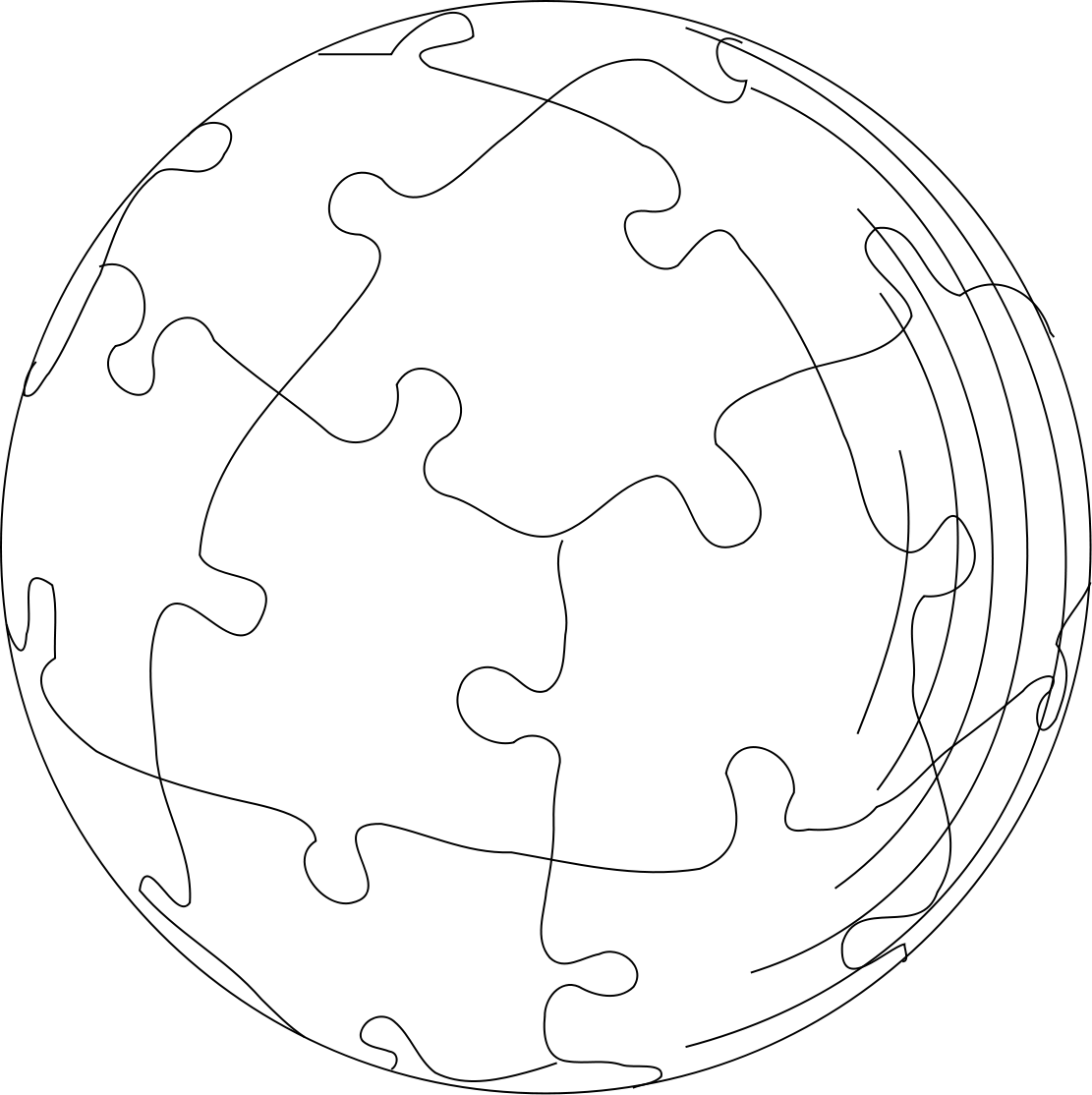}
\caption{\label{PuzzleBall} One way of thinking about the foundations of physical theories would have it that some of the central principles of a theory have a distinguished status as the ``truly fundamental'' principles.  An alternative view, which I describe and advocate here, is that the foundations of a theory are better thought of as a network of mutually interdependent principles, interlocking like the pieces of a spherical puzzle.  On this view, one would tend to expect that any of the central principles of a theory should be derivable from the rest of the theory with that principle removed, much like the overall shape of a puzzle ball constrains the shape of any individual piece.}
\end{figure}

It seems to me that these reflections suggest a proposal.  Instead of thinking of the foundations of a physical theory as consisting of a collection of essentially independent postulates from which the rest of the theory is derived, one might instead think of the foundations of a theory as consisting of a network of mutually interdependent principles---a collection of interlocking pieces, as in the spherical puzzle in figure \ref{PuzzleBall}.\footnote{\citet{Feynman} makes a distinction between two ways of understanding physical theories that is similar to the one I make here.  On the ``Greek'' view of theories, one begins with a collection of fixed fundamental axioms or postulates.  Feynman does not like this way of thinking about theories.  Instead, he endorses the ``Babylonian'' view, on which one observes that the principles of a theory are more richly connected: perhaps it is sometimes convenient to take certain principles of a theory as axioms and others as theorems, but one needs to recognize that in other cases one might want to switch this around and think of your theorems as the axioms, and use them to prove your former axioms.  He then observes that ``If all these various theorems are interconnected by reasoning there is no real way to say `These are the most fundamental axioms,' because if you were told something different instead you could also run the reasoning the other way. It is like a bridge with lots of members, and it is overconnected; if pieces have dropped out you can reconnect it another way'' (pg. 46).  The view I describe here is firmly in Feynman's Babylonian tradition.  I am grateful to Bill Wimsatt for pointing out this connection.}   The idea is that, as with the geodesic principle, one should generally expect that many of the central principles of a physical theory may be proved as theorems, given the rest of the theory.  Trying to make a distinction between the top-tier principles and the second-tier principles of a theory is not fruitful, then, since most, or even all, of the principles can be understood equally well as either postulate or theorem, and indeed, in different contexts it may well be desirable to think of them in different ways.  Importantly, theories are not modular, in the sense described above.  We cannot simply replace any given principle with some other one, at least not without changing the rest of the theory in possibly dramatic ways.  And theorems like the Geroch-Jang theorem and its Newtonian counterpart are of interest because they exhibit the details of these interdependencies.  They show just how the pieces interlock.

To be sure, nothing I have said thus far should count as an apodictic argument for the view I have described (call it the ``puzzle ball view'').  Nor will I give such an argument---indeed, I am not sure what an argument for the claim that there are no truly fundamental principles of general relativity would look like.  Instead, I merely offer the view as an alternative way to conceptualize the kinds of interrelations between the principles of physical theories on display with theorems like the Geroch-Jang theorem.  Perhaps the proposal is best conceived as a conjecture, albeit one with some compelling early evidence, for the following reason: while the senses in which, for instance, the conservation condition and the geodesic principle follow from other standard assumptions of general relativity are now established, the senses in which other principles, such as Einstein's equation or various energy conditions, are derivable from or constrained by the rest of the theory are less clear.  And so we have the skeleton of a mathematical question: are \emph{all} of the central principles of relativity theory and geometrized Newtonian gravitation (or other theories still) indeed mutually interderivable in the way that I have suggested?  Most or many of them?  Or are the geodesic principle and conservation condition anomalies?

Some important work has already been done on this topic: \citet{Dixon-GPE} has shown a sense in which the geometrized Poisson equation is the unique dynamical principle compatible with a collection of natural assumptions in Newtonian gravitation; similarly, \citet{Sachs+Wu} and \citet{Reyes} have shown that there is a sense in which the (vacuum form of) Einstein's equation can be derived from (in effect) the geodesic principle, among other assumptions, and \citet{Curiel} has argued that there is a sense in which the Einstein tensor is the unique tensor that can appear on the left hand side of Einstein's equation, even in the non-vacuum case.  Meanwhile, \citet{Duval+Kunzle} and \citet{Christian} have argued that even though the conservation condition in geometrized Newtonian gravitation does not follow from the geometrized Poisson equation, one can nonetheless derive it from other principles, at least if one considers Lagrangian formulations of the theory.  One might even understand Newton's argument for universal gravitation as a kind of heuristic argument that the inverse square law of gravitation is the unique dynamical principle compatible with certain other central principles of standard Newtonian gravitation (including generalized empirical facts, such as the universality of elliptical orbits).  Results such as these provide tentative evidence for my basic hypothesis, that one should expect many or all of the central principles of these spacetime theories to be mutually interdependent.

But I do not think these results are yet conclusive.  Specifically, what has not been done is to systematically study such results in order to try to characterize, in the way that has now been done for the geodesic principle theorems, (1) just what the assumptions going into these theorems are, (2) how natural the assumptions are in the contexts of the relevant theories, and (3) how these assumptions, in turn, depend on the other central principles of the theories (if they do).  And the attractiveness of the proposal presented here turns on the answers to these questions.  It is only after a project of this form has been carried out that one can fully evaluate whether the central principles of of these theories are really as tightly intertwined as the puzzle ball view would have it.  That said, if a careful study of this sort reveals that only some of the central principles of a theory are interconnected, it may still be fruitful to think about the foundations of theories in the way I propose here, since the discovery that some central principles of a theory (say, energy conditions in general relativity) are more peripheral than others need not imply that one can make sense of a unique or privileged collection of the most fundamental or basic principles.  Much will depend on just what the structure of the situation turns out to be.

It is worth emphasizing that mapping out these kinds of relations between the central principles of a physical theory is of some independent interest, since understanding the extent to which the central principles of general relativity in particular are mutually interdependent could play an important role in the construction of future theories (and in some ways, it already has) .\footnote{Feynman makes a related point about the practical importance of his Babylonian approach to theories. He writes, ``If you have a structure that is only partly accurate, and something is going to fail, then if you write it with just the right axioms maybe only one axiom fails and the rest remain, you need only change one little thing. But if you write it with another set of axioms they may all collapse, because they all lean on that one thing that fails. We cannot tell ahead of time, without some intuition, which is the best way to write it so that we can find out the new situation. We must always keep all the alternative ways of looking at a thing in our heads; so physicists do Babylonian mathematics, and pay but little attention to the precise reasoning from fixed axioms'' \citep[pg. 54\#]{Feynman}.}  The reason has to do with this idea of ``fiddling'' with physical theories.  There is a long-standing tradition of attempting to modify general relativity with small changes: for instance, in Brans-Dicke theory, one modifies Einstein's equation to include an additional scalar field; in TeVeS gravitational theories, one additionally considers vector fields.  In still other cases, one modifies general relativity by allowing derivative operators with torsion.  In each of these examples (and many others), one makes what appears to be a local change in the central principles of general relativity.

But these small changes can have dramatic consequences: for instance, in Einstein-Cartan theory, a modification of general relativity with torsion, the conservation condition does not generally hold.  One does not have a geodesic principle, at least in the ordinary sense, since in general the collection of self-parallel curves picked out by the derivative operator do not agree with the collection of extremal curves picked out by the metric, and free massive test point particles need not propagate along either class of curves.  Thus, apparently small tweaks can lead to a dramatically different theory, conceptually speaking.  A clearer picture of just how the central principles of general relativity \emph{do} fit together and constrain one another may provide important clarity into just what the ramifications of these ``small'' modifications to the theory are, and more, may help guide us in the search for alternative theories of gravitation, by indicating which principles are more or less tightly connected to which others.  Indeed, for this reason there is a sense in which the situation I described above, where some principles are very tightly interlocking and others turn out to be more loosely connected (for instance, some principles play a role as assumptions in some theorems, but cannot be proved in complete generality themselves) is the most interesting, from the practical perspective of mapping out the space of possible future theories.

In the next section, I will return to the question of explanation, now from the perspective of the present view.  But before I do so, I want to clarify the puzzle ball view slightly, as the language I have used to describe it may call to mind two other well-known ideas.  It seems to me that the view I have described is distinct from both.  First, note that the present proposal involves a picture of theories on which one emphasizes the ways in which the principles cohere with one another.  This way of thinking may be reminiscent of coherentism in epistemology, a variety of anti-foundationalism that holds that to justify a belief is to show how it coheres with one's other beliefs \citep[cf.][]{Kvanvig}.  But there is at least one major difference.  Coherentism takes the coherence of one's beliefs to be a form of justification for those beliefs.  Nothing about the puzzle ball view should be taken to suggest that the justification for general relativity comes from the apparent fact that one can derive certain central principles from others---rather, the justification for the theory is based on its empirical successes.  Or perhaps more precisely, our justification for general relativity is essentially independent of the relationship between the theory's central principles.  To see the point most clearly, one might well expect both general relativity and geometrized Newtonian gravitation to be coherent, in the sense of having mutually interdependent central principles. But this does not imply that they are equally well justified---indeed, general relativity is better justified than geometrized Newtonian gravitation even if the pieces of geometrized Newtonian gravitation are more tightly interlocking.\footnote{The suggestion of a connection to coherentism raises a second, related issue.  Even if we do not take the coherence of a body of beliefs as justification for any particular belief, one might nonetheless think of coherence as a virtue for a body of beliefs: all else being equal, one might tend to prefer to hold coherent beliefs than not.  Should one say the same thing about physical theories?  All else being equal, should one prefer as theory whose pieces interlock?  I am not sure that anything in the body of the paper depends on this, but I am inclined to say ``yes'', for several reasons.  First, as I argued above, when the central principles of theories are (partially) mutually interdependent, the theory provides a guide for the building of future related theories in a way that may be helpful for scientific practice.  Second, principles that are mutually interdependent are protected against claims of being \emph{ad hoc}.  A particular principle cannot be considered arbitrary or unmotivated if it is derivable, perhaps in multiple ways, from one's other principles.  To put this point in a more experimentally-oriented way, if the pieces of a theory are mutually interdependent, then testing any one principle can be understood as an implicit test of the other principles of a theory \citep[See][]{Harper}.  A third reason comes from Bill \citet{Wimsatt}, who argues that interderivability (or rather, multiple interderivability) is an indication of theoretic robustness and confers a kind of stability under theory change.}

Another view that the puzzle ball view may be reminiscent of is some variety of Quinean holism \citep[cf.][]{Quine}.  Quine famously used the ``web of belief'' metaphor when arguing for the interdependencies of our scientific beliefs, and against the analytic/synthetic distinction.  One might worry that the puzzle ball picture above is just an alternative metaphor used to make a strikingly similar point---indeed, the claim that we cannot make a fruitful distinction between top-tier and second-tier principles sounds like an argument against an analytic/synthetic distinction, at least in the narrow domain of the foundations of certain physical theories.  And perhaps it is right that I have recapitulated Quine here, though if it is, I think the point deserves to be made again since it is relevant for the present discussion of the geodesic principle.  Still, while this chapter is not the occasion for detailed Quine exegesis, I will point to two ways in what I have proposed is \emph{prima facie} different from Quine's holism, at least on the web-of-belief version.\footnote{My goal in the text is to distinguish the puzzle ball view from web of belief holism.  But this should not be taken to imply that Quine does not come much closer to the puzzle ball view in other parts of his opus.  For instance, \citet[Sec. V]{QuineCLT} distinguishes ``legislative postulates'' from ``discursive postulates''.  ``Legislative postulation,'' he writes, ``institutes truth by convention...'' whereas ``...discursive postulation is mere selection, from some pre\"existing body of truths, of certain ones for use as a basis from which to derive others, initially known or unknown'' \citep[pg. 360]{QuineCLT}.  He then goes on to argue that ``conventionality is a passing trait, significant at the moving front of science but useless in classifying the sentences behind the lines.  It is a trait of events and not of sentences.''  In other words, one might, when first developing a new scientific theory, begin with some bare, legislative postulates.  But as the theory develops, these truths ``...become integral to the corpus of truths; the artificiality of their origin does not linger as a localized quality, but suffuses the corpus.  If a subsequent expositor singles out those once legislatively postulated truths again as postulates, this signifies nothing; he is engaged only in discursive postulation.  He could as well choose his postulates from elsewhere in the corpus, and will if he think this serves his expository ends'' \citep[pg. 362]{QuineCLT}.  The idea, I take it, is that once one has a well-developed scientific theory---such as general relativity---one often identifies postulates for the purposes of deriving new facts about the theory, but these are always discursive, and more, which facts or statements of the theory one will take to be the postulates in any given case will depend on one's purposes.  This picture seems quite close to the puzzle ball view, indeed.  I am grateful to Pen Maddy for pointing out this connection to me.}

The first difference concerns just what the holism is supposed to be doing. Quine uses the interdependencies between beliefs as an argument for a radical form of conventionalism: when faced with evidence that conflicts with our beliefs, we have considerable leeway in choosing which parts of the web of beliefs to revise.  Indeed, the web image is supposed to support a distinction between ``central'' or ``core'' beliefs and ``peripheral'' beliefs such that we can always accommodate challenges to our full collection of beliefs by modifying only the peripheral beliefs and leaving the core beliefs intact.  But this is precisely the opposite of what I have argued here, at least with regard to the foundations of spacetime theories.  Instead, the idea is supposed to be that the foundations of physical theories are \emph{not} modular, and that in general one has remarkably little latitude in how one revises a theory in light of new evidence.  And this, I take it, is a desirable feature, since it provides a way out of the radical conventionalism I just described.  Since the various principles of a physical theory constrain one another, we have very few degrees of freedom for enacting minor changes in theories in light of new evidence.

The second difference is related (and relates, too, to coherentism as described above).  Quine's web of belief is supposed to be a (descriptive) metaphor for the sum total of one's beliefs.  The view I have described here is much narrower in its scope.  I do not claim that all of one's beliefs interlock in the way described; nor do I claim that scientific knowledge as a whole can be characterized by a puzzle ball.  The view does not even hold that particular scientific theories have this feature.  The suggestion is that the \emph{central principles} of some scientific theories are mutually interderivable, that the foundations of some physical theories should be though of in a certain way.  I have been deliberately vague about just what is supposed to count as a central principle, in large part because I think that trying to list these principles in advance, even for well understood theories such as general relativity or geometrized Newtonian gravitation, would be unproductive.  In fact, one might expect that a full account of just what the central principles of a theory are may have to wait until one sees just what assumptions are necessary nodes when trying to map out the network of interconnected principles at the heart of a given physical theory.  What I have done so far---and what I think can be done at this stage---is give examples of central principles of particular theories.  And so one can say that among the central principles of general relativity, for instance, are things such as the conservation condition, Einstein's equation, the geodesic principle, and various energy conditions.  But the point is that a claim about a collection of principles of this specific character is quite different from a claim about human knowledge quite broadly.

Note that this last point means that there is still a robust sense in which one can think of some parts of a theory as having a special ``fundamental'' status, even on the puzzle ball view.  Specifically, one might take \emph{all} of the central principles of a theory to be fundamental.  This leaves quite a bit of a theory as non-fundamental---for instance, particular predictions of a theory would not be among the central principles, and so these would not count as fundamental.  If the puzzle ball view is to be viewed as anti-foundational, then, it is only with regard to determinations of relative fundamentality \emph{among} the central principles of a theory.\footnote{Feynman, and \citet{Wimsatt}, argue that in cases where some principles can be proved in many different ways and others cannot be proved or can be proved from fewer starting cases, one can recover a different sense of ``fundamental'' principles, namely that the principles that can be proved in the most different ways should be understood as the most fundamental.  Note that this turns the idea discussed above---where the most fundamental principles were the top-tier principles from which other principles would be derived, not the ones most often derived themselves---on its head.  This idea is intriguing, but I mention it only to set it aside as it plays no role in the present discussion.}

\section{Explaining inertial motion, redux}

Now that I have described the puzzle ball view in some detail, I can return to the question of principle interest in this paper: namely, is there a sense in which we should understand the Geroch-Jang theorem and its Newtonian counterpart as explanations?  As a first remark, let me reiterate that if we are thinking in terms of the puzzle ball view, it does not make sense to think of theories in terms of ``top-tier'' principles and other, derived principles: in short, there is no way to make the distinction, at least among the central principles of the theory. None of the assumptions of a theory are distinguished as the truly basic or fundamental ones.  And if if this is right, then the kind of explanation that we apparently cannot get of the geodesic principle in general relativity and in geometrized Newtonian gravitation is uninteresting.  No, we cannot derive the geodesic principle in either theory from  more fundamental principles, but that is because it does not make sense of talk of unambiguously ``more fundamental'' principles in the first place.

Instead, what we can do is show how the geodesic principle in both of these theories fits into the rest of the puzzle (as it were).  This, too, may be understood as an answer to the question, ``Why do bodies move in the particular way that they do in the absence of an external force?''  These theorems reveal that in the absence of an external force, in the context of their respective theories, bodies \emph{must} move along timelike geodesics.  In other words, the other basic assumptions of the theory constrain the motion of (small) bodies.  Why timelike geodesic motion rather than any other?  Because in general relativity, we understand matter to be conserved, and to be such that observers always attribute instantaneous subluminal velocities to it at any point.  And it turns out that these assumptions, in the presence of the rest of the theory, imply that the only curves along which free massive test point particles can propagate are timelike geodesics.  If we are committed to the rest of general relativity, then there is only one candidate principle for inertial motion.

So do general relativity and/or geometrized Newtonian gravitation explain inertial motion?  Given the considerations just mentioned, I think the answer in both cases is ``yes'', so long as one understands ``explain'' in the right way.  At very least, these theorems provide deep insight and understanding into why bodies move in the particular way that they do in the absence of any external force---which is precisely what we were after when we asked the question. Moreover, the insight provided is that, in the context of the other central principles of the theories, the geodesic principle is \emph{necessary}, the only principle governing inertial motion that is compatible with our other principles.  It is a demonstration of precisely the ways in which the working parts of general relativity and geometrized Newtonian gravitation constrain one another.

That said, this kind of explanation differs in some important ways from other explanations that one may be accustomed to thinking about.  In particular, if the puzzle ball view is correct, the kind of explanation I have just described need not be asymmetrical.  That is, if general relativity might be said to explain inertial motion in the present sense by appealing to the fact that one can derive the geodesic principle from various other assumptions in the theory, one should \emph{not} conclude that the geodesic principle cannot play a role in other derivations that should also count as explanatory---even derivations of the assumptions going into the Geroch-Jang theorem or its Newtonian counterpart.  Indeed, one should expect that just as the geodesic principle is constrained by the other central assumptions of general relativity, so too are the conservation condition, the strengthened dominant energy condition, and even Einstein's equation constrained.  And by the same reasons I have offered above for the view that one might justly call the Geroch-Jang theorem an explanation of inertial motion, one might also say that explanations can be given for the conservation condition or Einstein's equation, by showing how these principles of general relativity are derivable from the other central principles of the theory.  In other words, general relativity explains inertial motion by appeal to Einstein's equation, but it may equally well explains Einstein's equation by appeal to the geodesic principle and other central assumptions of general relativity.

This observation may give some readers pause.  There is, by now, a long tradition of philosophers of science worrying about the so-called ``problem of explanatory asymmetry'' \citep[cf.][]{Bromberger}: intuitively, explanations appear to run in one direction and only one direction.  The trajectory of a comet may explain why we see a bright light in the nighttime sky once every few hundred years, but a bright light in the nighttime sky cannot explain the trajectory of a comet; the height of a flagpole may explain the length of its shadow at sunset, but the length of the shadow does not explain the height of the flagpole.  And so, many philosophers have argued, an account of explanation that allows symmetrical explanations---situations where A explains B and B explains A---is \emph{prima facie} unacceptable.

A few remarks are in order.  First, \citet{vanFraassen} has argued, I think correctly, that explanation should be understood as essentially pragmatic---and in particular, that explanations should only be understood as responses to certain classes of question.  To determine whether or not some particular response to a why question (say) should count as a satisfactory explanation depends on the context in which the question was asked and the particular demands of the questioner.  While in some contexts we might want to say that a particular explanation runs in only one direction, there may well be other contexts in which the explanation would run in the other direction.  If one is thinking in this way then the present example is simply a special case: if the question ``why do bodies move in the particular way that they do in the absence of an external force?'' is understood as ``does general relativity require us to adopt the geodesic principle as the central principle governing inertial motion?'', then one is rightly satisfied by a response along the lines of the Geroch-Jang theorem, even if in other contexts---i.e., in response to other questions---one might appeal to the geodesic principle to explain (say) Einstein's equation.

But there is also a more important point to make, here: I do not claim to be offering an ``account of explanation'', or anything like it.  I have not suggested that a necessary or even sufficient condition for being an explanation is to show how the thing to be explained ``fits in'' with the rest of a physical theory, in the sense that it is derivable from other central principles of a theory.  The point, rather, is to try to spell out the sense in which a particular class of theorems that show how the central principles of a spacetime theory fit together might be understood as explanatory---to say what, precisely, the theorems are doing, and why one might think of this as a kind of explanation, at least on the puzzle ball view.  Not all explanations work this way, nor do they need to in order for the story I have told here to be correct.  And so, the fact that in \emph{some} cases, we would want to say that if $A$ explains $B$ then $B$ cannot explain $A$ in no way undermines the claim that the present explanations simply do not work that way.

This point can be made most starkly by pointing to various other questions one can ask about inertial motion, even in general relativity, whose answers would be quite different from the Geroch-Jang theorem. Consider, for example, a question concerning a particular instance of inertial motion.  Why, one might ask, does the perihelion of Mercury's orbit precess?  One would answer this by appealing to some particular initial state of Mercury and features of spherically symmetric solutions to Einstein's equation to show that Mercury's orbit is the only allowed trajectory for a body with certain properties in a solar system like ours.  The geodesic principle may play a role in this argument, insofar as one might idealize Mercury as a free massive test point particle, and Einstein's equation may play a role, insofar as one would want to consider a spacetime that is a solution to the equation, but the argument would have nothing to do with the Geroch-Jang theorem.  And moreover, one would expect this sort of explanation to be asymmetric: Einstein's equation and the geodesic principle, along with some details concerning the state of the solar system and initial conditions for Mercury, explain the precession of the perihelion of Mercury; the precession of the perihelion of Mercury does not explain Einstein's equation or the geodesic principle.

But this is just the point.  If the Geroch-Jang theorem and its Newtonian counterpart should be countenanced as explanations, it is only because they are satisfactory answers to particular questions, and they are only explanatory in the context of \emph{those} demands for explanation.  A question concerning the orbit of Mercury is quite different from a question concerning the nature of inertial motion generally. And these theorems answer only the most general version of the question: why \emph{this} principle as opposed to any other?  This is no mean task, but it is a specific one, and it needs to be treated with care.

\section*{Acknowledgements}

Versions of this work have been presented at CEA-Saclay (Paris), Bergisch University Wuppertal, University of Pittsburgh, University of Chicago, University of Texas at Austin, Columbia University, Brown University, New York University, University of Western Ontario (twice!), University of California--Berkeley, University of California--Irvine, Yale University, and at the `New Directions' conference in Washington, DC.  I am grateful to very helpful comments and discussion from all of these audiences, and particularly to (in no special order) John Manchak, Giovanni Valente, Craig Callender, Alexei Grinbaum, Harvey Brown, David Wallace, Chris Smeenk, Wayne Myrvold, Erik Curiel, Ryan Samaroo, John Norton, John Earman, Howard Stein, Mike Tamir, Bryan Roberts, Shelly Kagan, Sahotra Sarkar, Josh Dever, Bob Geroch, Bill Wimsatt, David Albert, Tim Maudlin, Sherri Roush, Josh Schechter, and Chris Hill.  Special thanks are due to David Malament, Jeff Barrett, Kyle Stanford, and Pen Maddy for many helpful discussions and comments on previous versions of this work.  Thank you, too, to Dennis Lehmkuhl for organizing the 2010 workshop on which this volume is based, and for editing the volume.

\singlespacing

\end{document}